\documentstyle[epsfig,12pt]{article}
\newsavebox{\PSLASH}
\sbox{\PSLASH}{$p$\hspace{-1.8mm}/}

\begin{document}
\title{\bf Evolution of Thick Shells in Curved Spacetimes}
\author{Sh. Khosravi$^{1}$\footnote{Email Address:
khosravi\_sh@saba.tmu.ac.ir},  S. Khakshournia$^{2}$ \footnote{Email
Address: skhakshour@aeoi.org.ir}, and R. Mansouri$^{3}$ \footnote{On
sabbatical leave from Department of Physics, Sharif University of
Technology, and Institute for Studies in Physics and
  Mathematics(IPM),Tehran.Email Address: mansouri@hep.physics.mcgill.ca}
  \\ \\
{\small $^{1}$Dept. of Physics, Faculty of Science, Tarbiat
Moa'lem Univ., Tehran, Iran}\\
{\small $^{2}$Nuclear Research Center, Atomic Energy Organization
of Iran, Tehran, Iran}\\
{\small $^{3}$Dept. of Physics, McGill University, Montreal QC,
 H3A 2T8, Canada}}
\date{}
\maketitle
\begin{abstract}
We extend our previous formalism, in order to analyze the dynamics
of a general shell of matter with an arbitrary finite thickness
immersed in a curved spacetime. Within this new formulation we
obtain the equations of motion of a spherically symmetric dust thick
shell immersed in vacuum as well as Friedmann-Robertson-Walker
spacetimes.
\end{abstract}
\newpage
\section{Introduction}
Understanding localized matter distributions in astrophysics,
cosmology, and relativity has always been a matter of interest since
the beginning of general relativity. Recognizing the difficulty of
handling thick shells within relativity, early authors assumed the
idealization of a singular surface as a thin shell and tried to
formulate its dynamics in general relativity
\cite{Sen,Lancz,Darmois}. Einstein and Strauss used implicitly the
concept of a thick shell to embed a spherical star in a FRW universe
\cite{Einstein}. The new era of intense interest in thin shells and
walls began with the development of ideas relating to the phase
transitions in early universe and the formation of topological
defects. Again, mainly because of technical difficulties, strings
and domain walls were assumed to be infinitesimally thin
\cite{Vilenkin}. Interest in thin shells, or hypersurfaces of
discontinuity, thereafter received an impetus from cosmology of
early universe. The formulation of dynamics of such singular
surfaces \cite{Sen,Lancz,Darmois} was summed up in the modern
terminology by Israel \cite{Israel}. In the
Sen-Lanczos-Darmois-Israel (SLDI) formalism, thin shells are
regarded as idealized zero thickness objects, with a
$\delta$-function singularity in their energy-momentum and
Einstein tensors.\\
In contrast to thin walls, thickness causes new subtleties,
depending on how the thickness is defined and handled. Early
attempts to formulate thickness, mainly motivated by the outcome
of the idea of late phase transitions in cosmology \cite{Hill},
were concentrated on domain walls. Silveira \cite{Silveria}
studied the dynamics of a spherical thick domain wall by
appropriately defining an average radius $<R>$, and then used the
well-known plane wall scalar field solution as the first
approximation to derive a formula relating $<\ddot{R}>$,
$<\dot{R}>$, and $<R>$ as the equation of motion for the thick
wall. Widrow \cite{Widrow} used the Einstein-scalar equations for
a static thick domain wall with planar symmetry. He then took the
zero-thickness limit of his solution and showed that the
orthogonal components of the energy-momentum tensor would vanish
in that limit. Garfinkle and Gregory \cite{Garfinkle} presented a
modification of the Israel thin shell equations to treat the
evolution of thick domain walls in vacuum. They used an expansion
of the coupled Einstein-scalar field equations describing the
thick gravitating wall in powers of the thickness of the domain
wall around the well-known solution of the hyperbolic tangent kink
for a $\lambda\phi^{4}$ wall. They concluded that the effect of
thickness in the first approximation was to effectively reduce the
energy density of the wall compared to the thin case, leading to a
faster collapse of a spherical wall in vacuum (see also
\cite{Bonjour}). Barrab\'{e}s \textit{et al} \cite{Barrabes}
applied the expansion to the wall action and integrated it
perpendicular to the wall to show that the effective action for a
thick domain wall in vacuum, apart from the usual Nambu term,
consists of a
contribution proportional to the induced Ricci curvature scalar.\\
Renewed interest in walls and branes, specially in higher
dimensions, came from string theory, mainly because it provided a
novel approach for resolving the cosmological constant and the
hierarchy problems \cite{Hamed}. In this scenario gravitation is
localized on a brane reproducing effectively four dimensional
gravity at large distances due to the warp geometry of the spacetime
\cite{Randall}. There are a large number of papers in last years
dealing with different properties of thick walls and branes. Most of
these papers use a solution of Einstein-scalar field equation in
n-dimensional spacetime representing a codimension-1 or 2 brane, or
a smoothing or smearing mechanism to modify the Randall-Sundrum
ansatz  \cite{group}. However, as far as the geometry of the problem
is concerned- irrespective of the spacetime dimension and the
motivation for considering a wall or brane- these models are all
based on a regular solution of Einstein equations on a manifold with
specified asymptotic behaviour representing a localized scalar
field, as it was assumed
in early works \cite{Goetz}.\\
A completely different method based on the gluing of a thick wall,
considered as a regular manifold, to two different manifolds on both
sides of it, was suggested in
 \cite{Mansouri1}. The idea behind this suggestion is to understand
the dynamics of a localized matter distribution of any kind confined
within two principally different spacetimes or matter phases. Such a
matching of three different manifolds appears to have many
applications in astrophysics, early universe, and string cosmology.
It enables one to have any topology and any spacetime on each side
of the thick wall or brane. One may apply it to the dynamics of
galaxy clusters and their halos or to a brane in any spacetime
dimension with any symmetry on each side of it. By definition, such
a matching is regular and there is no singular surface in this
formulation. Therefore {\it Darmois junction conditions} for the
extrinsic curvature tensor on the thick shell boundaries with the
two embedding spacetimes can be applied. They have applied their
definition of a thick wall to the special case of a spherically
symmetric dust thick shell with constant coordinate thickness. Using
an expansion scheme in the thickness of the shell they were then
able to obtain an approximate equation of motion for the thick
shell in vacuum.\\
Our aim is to generalize their formalism and give the basic
dynamical equations governing a thick codimension-1 brane, relaxing
some of their assumptions \cite{Mansouri1}. Specifically, we make no
assumption of symmetry or the equation of state of the brane matter,
and introduce the proper thickness of the wall, instead of using a
comoving wall thickness. We then apply our formalism to a dust thick
shell immersed in vacuum and also in a
Friedmann-Robertson-Walker(FRW) spacetime, taking into account the
peculiar velocity of the dust brane layers with respect to the
background geometry, in contrast to the previous
work where it was neglected \cite{Mansouri1}.\\
The paper is organized as follows. In section 2 we develop a thick
wall formalism yielding a generic equation to study the evolution of
thick shells in curved spacetime. In section 3, we apply our
formalism to obtain the equations of motion of a spherical thick
shell of dust matter in vacuum. Section 4 deals with a thick wall
in a FRW spacetime. We conclude then in section 5.\\
Natural geometrized units, in which $G=c=1$ are used throughout the
paper. The two boundary limits of the thick wall are called
$\Sigma_j$ with $j=1,2$. The core of the thick wall is denoted by
$\Sigma_0$. For any quantity $S$ let $S_{0}$ denote
$S|_{\Sigma_{0}}$.  Square bracket $[F]$ indicates the jump of any
quantity $F$ across $\Sigma_{j}$. Latin indices range over the
intrinsic coordinates of $\Sigma_{j}$ denoted by $\xi^{a}_{j}$, and
Greek indices over the coordinates of the 4-manifolds.

\section{Modelling the thick wall and its junction conditions}

The technology of manipulating thin and thick walls in general
relativity in arbitrary dimension are basically different. Thin
walls can be treated in two different but equivalent ways. Either
one solves Einstein equations in $d+1$ dimension with a {\it
distributional} energy momentum tensor which mimics an
infinitesimally thin wall carrying some kind of matter( dark or not
dark, including radiation), or one takes known solutions of the
Einstein equations on every side of the wall and glues them to the
wall by applying boundary conditions at the wall location. The
equivalence of these two procedures is not trivial but has been
shown rigorously for the general case \cite{Mansouri2}. Boundary
surfaces without any energy-momentum tensor can just be considered
as a special case. It should be noted that such an equivalence does
not exist for the codimension-2 walls or defects like strings in 4
spacetime dimension. The lack of such an equivalence in the case of
thick walls forces us to differentiate
between various definitions of thick walls or branes. \\
Usually for a thick wall, one takes a solution of the Einstein
equations representing a localized scalar field having a
well-defined asymptotic behaviour \cite{Garfinkle,Bonjour}. Although
this maybe a reasonable assumption, its application is however
limited. We prefer to accept another definition, first formalized in
\cite{Mansouri1}, being more suitable for different applications in
astrophysics and string cosmology. However, it must be mentioned
that a similar approach has been used in \cite{Katz} for the special
case of a spherically symmetric thick shell along with an
application to a more restricted junction condition in a two-fluid
model of oscillations of a neutron star \cite{Andersson}.
\\Now assume a localized matter distribution to be considered as a
solution of Einstein equations on a specific manifold with
well-defined boundaries. This 'thick wall' is then immersed in a
universe which is defined in principle differently on different
sides of the wall, being infinite in the planar or cylindrical case
or compact in the spherical case. Therefore, in contrast to the thin
wall formalism where one glues two different manifolds along a
singular hypersurface, our definition of a thick wall leads to the
problem of gluing three
different manifolds along non-singular hypersurfaces.\\
Consider a thick wall with two boundaries $\Sigma_{1}$ and
$\Sigma_{2}$ dividing the space-time manifold $\cal M$ into three
regions: ${\cal M}_{-}$ and ${\cal M}_{+}$, both outside the wall,
and ${\cal M}_{w }$ within the thick wall itself such that $\partial
{\cal M}_{-}\cap\partial {\cal M}_{w}=\Sigma_{1}$ and $\partial
{\cal M}_{w}\cap\partial {\cal M}_{+}=\Sigma_{2}$. Treating the two
surface boundaries $\Sigma_1$ and $\Sigma_2$, which separate the
manifold ${\cal M}_w$ from two distinct manifolds ${\cal M}_{-}$ and
${\cal M}_{+}$ respectively, as nonsingular timelike hypersurfaces,
is equivalent to the continuity of the intrinsic metric $h_{ab}$ and
the extrinsic curvature tensor $K_{ab}$ of $\Sigma_j\hspace{0.2cm}
,(j=1,2)$
 across the corresponding hypersurfaces. In other words, there
is no distribution of energy-momentum concentrated on the thick wall
boundaries $\Sigma_j$. These crucial requirements have been first
formulated in \cite{Mansouri1}:
\newpage
\begin{equation}\label{hmn}
[h_{ab}]\stackrel{\Sigma_{j}}{=}0,\quad\quad\quad\quad \quad \quad
j=1,2,
\end{equation}
\begin{equation}\label{kmn}
[K_{ab}]  \stackrel{\Sigma_{j}}{=} 0, \quad\quad\quad\quad
\quad\quad j=1,2.
\end{equation}
Note that in principle the boundaries $\Sigma$ could also be
space-like. Our restriction to time-like ones is just to simplify
the formalism as much as possible and if required, our formalism
maybe trivially applied to space-like cases. Now if we write
(\ref{kmn}) on $\Sigma_{1}$ and $\Sigma_{2}$ separately, and add
these two equations, we obtain
\begin{equation}\label{kmn2}
K_{ab} \Bigl|^{+}_{\Sigma_{2}} -K_{ab}
\Bigl|^{-}_{\Sigma_{1}}+K_{ab} \Bigl|^{w}_{\Sigma_{1}}- K_{ab}
\Bigl|^{w}_{\Sigma_{2}}=0,
\end{equation}
where the superscripts $-(+)$ and $w$ mean that the extrinsic
curvature tensor $K_{ab}$ of $\Sigma_{1}(\Sigma_{2}$) is evaluated
in the regions ${\cal M}_{-}({\cal M}_{+})$ and ${\cal M}_{w}$,
respectively. For convenience, we introduce a Gaussian normal
coordinate system $(n, \xi^{a}_{0})$ in the neighborhood of the core
of the thick shell denoted by $\Sigma_{0}$, corresponding to $n=0$.
$\xi^{a}_{0}$ are the intrinsic coordinates of $\Sigma_{0}$, and $n$
is the proper length along the geodesics orthogonal to $\Sigma_{0}$.
\par
Let us now expand the extrinsic curvature tensors in a Taylor series
around $\Sigma_{0}$ situated at $n=0$:
\begin{equation}\label{expansion}
K_{ab}\Bigl|_{\Sigma_{j}}=K_{ab}
\Bigl|_{\Sigma_{0}}+\varepsilon_{j}\, w\frac{\partial
K_{ab}}{\partial n} \Bigl|_{\Sigma_{0}} +o(w^2),
\end{equation}
where $2w$ denotes the proper thickness of the wall, and
$\varepsilon_{1}=-1$ and $\varepsilon_{2}=+1$. The above expansion
is justified if the proper thickness of the wall is small relative
to the radius of curvature (\ref{expansion}). Should this assumption
not be valid, we either have to go to the higher order terms, or to
use directly the exact equation (\ref{kmn2}). The derivative of the
extrinsic curvature is now given by
\begin{equation}\label{lie}
\frac{\partial K_{ab}}{\partial n} =K_{ac}K^{c}_{b}-R_{\mu\sigma
\nu\lambda}e^{\mu}_{a}e^{\nu}_{b}n^{\sigma}n^{\lambda},
\end{equation}
where $n^{\mu}$ is the normal vector field, and
$e^{\mu}_{a}=\partial x^{\mu}/\partial\xi^{a}_{0}$  are the three
basis vectors tangent to $\Sigma_{0}$. Substituting
(\ref{expansion}) into (\ref{kmn2}) and taking into account
(\ref{lie}), we arrive at :
\begin{eqnarray}
 K_{ab} \Bigl|^{+}_{\Sigma_{0}}-K_{ab} \Bigl|^{-}_{\Sigma_{0}}&+&
w\left((K_{ac}K^{c}_{b}-R_{\mu\sigma
\nu\lambda}e^{\mu}_{a}e^{\nu}_{b}n^{\sigma}n^{\lambda})
\Bigl|^{-}_{\Sigma_{0}} +(K_{ac}K^{c}_{b}-R_{\mu\sigma
\nu\lambda}e^{\mu}_{a}e^{\nu}_{b}n^{\sigma}n^{\lambda})
\Bigl|^{+}_{\Sigma_{0}}\right.\nonumber\\
&-&\left.2(K_{ac}K^{c}_{b}-R_{\mu\sigma\nu\lambda}
e^{\mu}_{a}e^{\nu}_{b}n^{\sigma}
n^{\lambda})\Bigl|^{w}_{\Sigma_{0}}\right)=0. \label{fund}
\end{eqnarray}
This is the basic equation for the dynamics of a thick wall in a
curved spacetime, written up to the first order in $w$. While the
metrics in $\cal M_{-}$ and $\cal M_{+}$ are usually given in
advance, knowing the metric in the wall spacetime ${\cal M}_{w}$,
however, requires a nontrivial work. Equation (\ref{fund}), or the
corresponding exact one (\ref{kmn2}), is to be considered as the
generalization of the Israel's thin wall condition \cite{Israel} to
the thick case. We will apply it now to particular examples of a
spherically symmetric dust thick shell in vacuum as well as to a FRW
universe in which $w$ is independent of angular coordinates.
However, (\ref{fund}) is valid for more general cases without
spherical symmetry where $w$ is not a constant. Furthermore, $w$ in
general may be a function of time but for the sake of simplicity
here we take it to be constant in time.
\section{Motion of a spherical thick dust shell in vacuum}

As the first example we consider a thick dust shell in vacuum. We
first derive the basic exact equations underlying the dynamics and
expand it in powers of the thickness, and then calculate the
peculiar velocity and its impact on the collapse of the shell
following by a comparison to the thin shell limit already known. We
constrain our calculations to the special case of motion of a shell
represented by an LTB solution where no shell crossing occurs
\cite{Bondi}.

\subsection{ Equations of motion}

Consider a spherically symmetric dust thick shell immersed in
vacuum. The spacetime exterior to the shell is Schwarzschild, and
the interior is taken to be the Minkowski flat spacetime:
\begin{eqnarray}\label{metric}
ds^{2}\Bigl|_{out}&= -f
dt_{+}^{2}+\textit{f}^{-1}dr_{+}^{2}+r_{+}^{2}(d\theta^{2}
+\sin^{2} \theta
d\varphi^{2}),\\
ds^{2}\Bigl|_{in}&= -dt_{-}^{2}+dr_{-}^{2}+r_{-}^{2}(d\theta^{2}
+\sin^{2} \theta d\varphi^{2}),
\end{eqnarray}
where $f=1-(2M/r_{+})$, $M$ being the mass of the thick shell. To
allow for the inhomogeneity of the mass distribution within the
shell, we take its metric to be that of the
Lemaitre-Tolman-Bondi(LTB). In the synchronous comoving coordinates
$(t , r, \theta , \varphi )$, the LTB metric takes the following
form \cite{Bondi}:
\begin{equation}\label{LTB}
ds^{2}\Bigl|_{w}=-d t^{2} +\frac{R'^{2}}{1+2E(r)} \,\,d r^{2}
+R^{2}(r, t)(d\theta^{2} +\sin^{2} \theta d\varphi^{2}),
\end{equation}
where the prime denotes the partial differentiation with respect to
$r$,  and $E(r)$ is an arbitrary real function such that
$E(r)>-(1/2)$ and $R'>0$ ( because of the no shell crossing
condition). The corresponding Einstein field equations then turn out
to be
\begin{equation}\label{Eins}
R_{,t}^{2} (t,r) =2E(r) +\frac{2M(r)}{R}\qquad ,\qquad  \rho (t,r) =
\frac{M'(r)}{R^{2}R'},
\end{equation}
where $\rho (t,r)$ is the energy density of the dust fluid in ${\cal
M}_{w}$, and $M(r)$ is another arbitrary real smooth
function interpreted as the mass.\\
Let us now proceed to apply (\ref{fund}) in order to investigate the
evolution of the spherical thick shell described above.  The angular
component of (\ref{fund}) for the thick shell in vacuum is given by
\begin{eqnarray}
K_{\theta\theta}\Bigl|^{out}_{\Sigma_{0}}-K_{\theta\theta}\Bigl|^{in}_{\Sigma_{0}}+
w\left(K_{\theta\theta}K^{\theta}_{\theta}\Bigl|^{in}_{\Sigma_{0}}
+(K_{\theta\theta}K^{\theta}_{\theta}-R_{\theta\sigma\theta\lambda}n^{\sigma}n^{\lambda})
\Bigl|^{out}_{\Sigma_{0}}
-\right.\nonumber\\
\left. 2(K_{\theta\theta}K^{\theta}_{\theta}-
R_{\theta\sigma\theta\lambda}n^{\sigma}n^{\lambda})\Bigl|^{w}_{\Sigma_{0}}\right)=0.
\label{kthth}
\end{eqnarray}
\newpage
The corresponding time component of (\ref{fund}) is
\begin{eqnarray}\label{eqktt}
K_{\tau\tau} \Bigl|^{out}_{\Sigma_{0}}-K_{\tau\tau}
\Bigl|^{in}_{\Sigma_{0}}+w\left(K_{\tau\tau}K^{\tau}_{\tau}\Bigl|^{in}_{\Sigma_{0}}+(K_{\tau\tau}K^{\tau}_{\tau}
-R_{\mu\sigma\nu\lambda}n^{\sigma}n^{\lambda}u^{\mu}u^{\nu})\Bigl|^{out}_{\Sigma_{0}}\right. -\nonumber\\
\left.2(K_{\tau\tau}K^{\tau}_{\tau}
-R_{\mu\sigma\nu\lambda}n^{\sigma}n^{\lambda}u^{\mu}u^{\nu})
\Bigl|^{w}_{\Sigma_{0}}\right) =0,
\end{eqnarray}
where $u^{\mu}$ is the four velocity tangent to $\Sigma_{0}$. Armed
with (\ref{kthth}) and (\ref{eqktt}), we are now in a position to
write down explicitly the equation of motion of the thick shell.
Note that the four velocity $u^{\mu}$ and the normal vector
$n^{\mu}$ on the spherical thick shell's core $\Sigma_{0}$ for the
Schwarzschild and LTB spacetimes are, respectively
\begin{eqnarray}\label{unout}
u^{\mu}\Bigl|^{out}_{\Sigma_{0}}&=&(\dot{t_{+}},\dot{R},0,0)\Bigl|_{\Sigma_{0}},\nonumber\\
n^{\mu}\Bigl|^{out}_{\Sigma_{0}}&=&(f^{-1}\dot{R},f\dot{t_{+}},0,0)\Bigl|_{\Sigma_{0}},\\
\label{unwall}
u^{\mu}\Bigl|^{w}_{\Sigma_{0}}&=&(\dot{t},\dot{r},0,0)\Bigl|_{\Sigma_{0}},\nonumber\\
n^{\mu}\Bigl|^{w}_{\Sigma_{0}}&=&\left(v_{0} \dot{t},
\frac{\sqrt{1+2E}}{R'}\dot{t},0,0\right)\Bigl|_{\Sigma_{0}},
\end{eqnarray}
where the dot denotes the derivative with respect to the proper time
$\tau_{0}$ on $\Sigma_{0}$, and $v_{0}$ is the peculiar velocity of
$\Sigma_{0}$ relative to the LTB geometry defined as
\begin{equation}\label{pecvel}
v_{0}=\frac{R'}{\sqrt{1+2E}}\frac{d r}{d t}|_{\Sigma_{0}},
\end{equation}
and is related to the Lorentz factor in the LTB geometry as
\begin{equation}\label{tdot}
\dot{t}|_{\Sigma_{0}}=\frac{1}{\sqrt{1-v_{0}^{2}}}\,\,.
\end{equation}
Note that the definition of the peculiar velocity, (\ref{pecvel}),
ceases to be valid for a thin shell, in which $d r/d t=0$. Relevant
components of the Riemannian curvature tensor for the Schwarzschild
spacetime are
\begin{equation}\label{Riemn}
R_{\theta rr\theta}=\frac{M}{fr_{+}}\, ,\hspace{0.4cm}
R_{trrt}=\frac{2M}{r_{+}^3}\, ,\hspace{0.4cm} R_{\theta tt
\theta}=-\frac{{M}f}{r_{+}}\, ,
\end{equation}
and for the LTB spacetime are
\begin{eqnarray}\label{Riemn2}
R_{\theta rr\theta} &=&
\frac{-RR'}{(1+2E)}\,(R_{,t}R'_{,t}-E')\, \, ,\nonumber\\
R_{trrt}&=&\frac{R'}
{1+2E}R'_{,tt}\,\,\, ,\nonumber\\
R_{\theta tt\theta} &=& RR_{,tt}\,\,\,.
\end{eqnarray}
The components of the extrinsic curvature tensor on $\Sigma_{0}$
evaluated with respect to the relevant regions are given by
\begin{eqnarray}\label{kthth1}
K_{\theta}^{\theta}\Bigl|_{\Sigma_{0}}^{in}&=&\frac{1}{R_{0}}\sqrt{1+\dot{R_{0}}^2}\,\, ,\nonumber\\
K_{\theta}^{\theta}\Bigl|_{\Sigma_{0}}^{w}&=&\frac{1}{R_{0}}\sqrt{1+\dot{R_{0}}^2
-\frac{2M_{0}}{R_{0}}}\,\, ,\nonumber\\
K_{\theta}^{\theta}\Bigl|_{\Sigma_{0}}^{out}&=&
\frac{1}{R_{0}}\sqrt{1+\dot{R_{0}}^2-\frac{2M}{R_{0}}}\,\, ,
\end{eqnarray}
and
\begin{eqnarray}\label{ktt}
K_{\tau}^{\tau}\Bigl|^{in}_{\Sigma_{0}}&=&\frac{\ddot{R_{0}}}{\sqrt{1+\dot{R_{0}}^2}}\,\, ,\nonumber\\
K_{\tau}^{\tau}\Bigl|^{w}_{\Sigma_{0}}&=&\frac{\ddot{R_{0}}+\frac{M_{0}}{R_{0}^2}
+ R_{0}\frac{\rho_{0}v_{0}^{2}}{1-v_{0}^{2}}}{\sqrt{1+\dot{R_{0}}^2-
\frac{2M_{0}}{R_{0}}}}\,\, ,\nonumber\\
K_{\tau}^{\tau}\Bigl|^{out}_{\Sigma_{0}} &=&\frac{\ddot{R_{0}}
+\frac{M}{R_{0}^2}}{\sqrt{1+\dot{R_{0}}^2-\frac{2M }{R_{0}}}}\,\,.
\end{eqnarray}
Substituting (\ref{unout}-\ref{ktt}) into (\ref{kthth}) and
(\ref{eqktt}), after some manipulations we end up with two
independent equations written up to the first order of $w/R_{0}$ as
follows:
\begin{equation}\label{thickeq1}
\sqrt{1+\dot{R_{0}}^2}-\sqrt{1+\dot{R_{0}}^2-\frac{2M}{R_{0}}}=
\frac{8\pi w\rho_{0}R_{0}}{1-v_{0}^2}-w\frac{M-2M_{0}}{R_{0}^2} ,
\end{equation}
and
\begin{eqnarray}\label{thickeq2}
(\alpha-\beta)\ddot{R_{0}}=-\frac{\alpha M}{R_{0}^2}-
w\alpha\beta\left(\frac{\ddot{R}_{0}^2}{1+\dot{R}_{0}^2}
+\frac{(\ddot{R}_{0}+\frac{M}{R_{0}^2})^2}{1+\dot{R}_{0}^2-\frac{2M}{R_{0}}}
-\frac{2(M-2M_{0})}{R_{0}^3}-\right.\nonumber\\
\left.\frac{2(\ddot{R}_{0}+\frac{M_{0}}{R_{0}^2}+4\pi
R_{0}\frac{\rho_{0}v_{0}^2}{1-v_{0}^2})^2}{1+\dot{R}_{0}^2-\frac{2M_{0}}{R_{0}}}
-8\pi \rho_{0}\right),
\end{eqnarray}
where for convenience we have defined
\begin{equation}\label{definition}
\alpha=\sqrt{1+\dot{R_{0}}^2},\hspace{1cm}
\beta=\sqrt{1+\dot{R_{0}}^2-\frac{2M}{R_{0}}}.
\end{equation}
The peculiar velocity $v_0$ can also be written in an explicit form.
The angular component of the extrinsic curvature of $\Sigma_{0}$ in
the Gaussian normal coordinate is
\begin{equation}\label{kthth2}
\hspace{1cm} K^{\theta}_{\theta}=\frac{1}{R}\frac{\partial
R}{\partial n}\Bigl|_{\Sigma_{0}} =\frac{1}{R}\left(\frac{\partial
R}{\partial t}\frac{\partial t}{\partial n}+\frac{\partial
R}{\partial r}\frac{\partial r}{\partial
n}\right)\Bigl|_{\Sigma_{0}} =\frac{\dot{t}}{R}(v
R_{\,,t}+\sqrt{1+2E})\Bigl|_{\Sigma_{0}},
\end{equation}
where we have used (\ref{unwall}). On the other hand one can write
\begin{equation}\label{DotR}
\dot{R}=\dot{t}(R_{\,,t}+v\sqrt{1+2E})\Bigl|_{\Sigma_{0}}.
\end{equation}
Eliminating $\dot{t}$ from (\ref{kthth2}) and (\ref{DotR}) and using
the expression (\ref{kthth1}) for $K^{\theta}_{\theta}$ evaluated
with respect to LTB geometry we deduce the following formula
\begin{equation}\label{upsilon}
v_{0}=\frac{\beta\sqrt{2E_{0}+\frac{2M_{0}}{R_{0}}}-
\dot{R_{0}}\sqrt{1+2E_{0}}}{\dot{R_{0}}\sqrt{2E_{0}+\frac{2M_{0}}{R_{0}}}
-\beta\sqrt{1+2E_{0}}}\,\, ,
\end{equation}
 which is valid only for the shells of finite thickness.
Equations (\ref{thickeq1}) and (\ref{thickeq2}) together with
(\ref{upsilon}) are the dust thick shell equations of motion written
up to the first order in terms of the shell's proper thickness.
Given the initial data $R_{0}(0)$, $\dot{R}_{0}(0)$, $E_{0}$,
$M_{0}$, and $w$, one can in principle solve this set of
differential equations to determine the time evolutions of the
proper radius $R_{0}(\tau_{0})$ and and
the mass density $\rho_{0}$ of the thick shell's core.\\

\subsection{Impact of thickness and the thin wall limit}

In order to investigate the impact of the proper thickness as well
as the peculiar velocity on the dynamics of the shell we take a
closer look at (\ref{thickeq1}). For the sake of comparison to the
thin wall case, we write down the following relation between the
surface energy density of the dust thin shell and the energy density
of the dust thick shell evaluated on $\Sigma_{0}$ up to the first
order in $w$:
\begin{equation}\label{rho}
\sigma=\int_{-w}^{w} \rho\, dn\simeq 2w\rho_{0}+o(w^2).
\end{equation}
Now, assuming $\upsilon_{0}\ll1$, the (\ref{thickeq1}) can be
written as:
\begin{equation}\label{Equation}
\alpha-\beta=4\pi \tilde{\sigma}R_{0},
\end{equation}
where we have singled out an effective surface density
$\tilde{\sigma}$ defined by
\begin{equation}\label{sigma}
\tilde{\sigma}=\sigma\left(1+\upsilon_{0}^{2}-w\frac{M-2M_{0}}{4\pi\sigma
R_{0}^3}\right).
\end{equation}
Equation (\ref{Equation}) has the well-known form of the dynamics of
a spherical thin shell with the effective surface
energy density $\tilde{\sigma}$ \cite{Israel,Mansouri3}.\\
The zero thickness limit of the shell is therefore defined by
$w\rightarrow0$ and $v_{0}\rightarrow0$, as mentioned in the
paragraph following (\ref{tdot}). Note that in this limit the
density $\rho$ blows up while the integral (\ref{rho}) remains
fixed, so that all the information of internal structure of the
shell is squeezed in the surface energy density $\sigma$. Taking
this limit, (\ref{Equation}) reduces to the familiar equation of
motion for the dust thin shell in vacuum \cite{Israel,Mansouri3}.
The case of vanishing peculiar velocity, i.e. if the shell's core is
comoving with the LTB expansion $(\upsilon_{0}=0)$, is specially
easy to discuss. Assuming $M > 2M_{0}$, which is usually the case,
we have $\tilde{\sigma}<\sigma$. Solving (\ref{Equation}) for
$\dot{R}^{2}$, we get
\begin{equation}\label{Rdot}
\dot{R}_{0}^{2} =4\pi^{2}  R_{0}^{2} \tilde{\sigma}^{2}
+\frac{{M}^{2}}{16\pi^{2}R_{0}^{4} \tilde{\sigma}^{2}}
+\frac{{M}}{R_{0}} -1.
\end{equation}
We now focus our attention on the dependence of $\dot{R}_{0}^{2}$ on
$\tilde{\sigma}$ for a given shell's radius $R_{0}$ and a given
initial value for $M$ to see how the thickness affects the collapse
velocity. Taking into account $R_{0}>2M$, which means roughly that
the radius of the shell's core is greater than its Schwarzschild
radius, we obtain for the zeros of $\dot{R}_{0}^{2}$ (Figure 1):
\begin{figure}[t]
\begin{picture}(150,135)(-100,30)
\epsfig{file=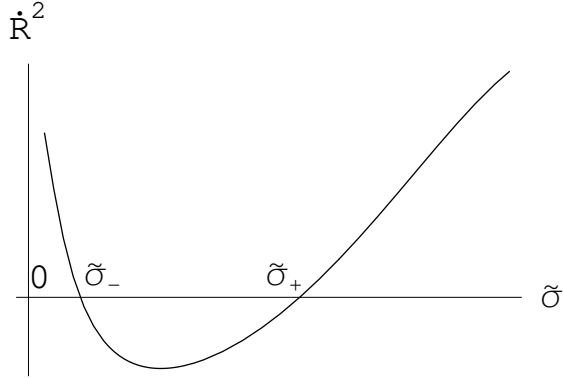}
\end{picture}
\vskip 30pt
 \caption{Schematic plot of $\dot{R}^2$ versus
$\tilde{\sigma}$}
\end{figure}
\begin{equation}\label{Zeroes}
\tilde{\sigma}_{\pm}^{2}=\frac{1 \pm \frac{M}{R_0}-
\sqrt{1-\frac{2M}{R_0}}}{8\pi^{2} R_{0}^{2}}.
\end{equation}
The branch $\tilde{\sigma}_{-}<\tilde{\sigma} <\tilde{\sigma}_{+}$,
leading to $\dot{R}_{0}^{2}< 0$, is not physical. The branch
$\tilde{\sigma}>\tilde{\sigma}_{+}$ corresponds to the equation
 $\alpha+\beta=4\pi \tilde{\sigma}R$, arising from
the process of squaring (\ref{Equation}), turns out to be irrelevant
to our problem, as we are just considering the ordinary centered
shells (black hole type matching) \cite{Mansouri4}. We are left,
therefore, with the branch corresponding to
$\tilde{\sigma}<\tilde{\sigma}_{-}$ in which $\dot{R}_{0}^{2}$ is a
decreasing function of $\tilde{\sigma}$. Hence, for a comoving
$\Sigma_{0}$ $(\upsilon_{0}=0)$, having $\tilde{\sigma}<\sigma$, the
thickness  in the first order leads to a faster collapse of the
thick shell relative to a thin shell in vacuum. This corresponds to
the case discussed in \cite{Mansouri1}. While for a $\Sigma_{0}$
with a small peculiar velocity with respect to the LTB background,
the effective energy density $\tilde{\sigma}$ tends to increase
leading to a slowdown of the collapse velocity
 of the shell relative to a comoving one.

\section{Motion of a spherical dust thick shell in
a FRW spacetime}

Dust thick shells embedded in FRW space times may have different
applications. The dynamics of great wall in astrophysics,
 supernovae envelopes, and voids are just few examples. It is different
 from the vacuum case as the space time is not any
 more asymptotically Minkowski. The dynamics of FRW in and outside of
 the shell makes it necessary to take care of the
 peculiar velocity relative to both FRW space times which make the calculation
lengthier.

\subsection{Equations of motion}

Consider a spherically symmetric thick shell of dust matter
separated by two different dust-filled FRW spacetimes whose metrics
are given by
\begin{eqnarray}\label{FRW1}
d s^{2}\Bigl|_{in}&=-d t_{in}^{2} +\textit{a}_{\,
in}(t_{in})^2\left( d\chi_{in}^{2}
+r_{in}(\chi_{in})^{2}(d\theta^{2}
+\sin^{2} \theta d\varphi^{2})\right), \\
\label{FRW2} d s^{2}\Bigl|_{out}&=-d t_{out}^{2} +\textit{a}_{\,
out}(t_{out})^2\left( d\chi_{out}^{2}
+r_{out}(\chi_{out})^{2}(d\theta^{2} +\sin^{2} \theta
d\varphi^{2})\right),
\end{eqnarray}
where
\begin{eqnarray}\label{defFRW}
r(\chi)=\left\{ \begin{array}{ll}\sin \chi &(k=+1,
\hspace*{0.55cm}\mbox{closed universe}),\\
\chi
&(k=0,\hspace*{0.55cm}  \mbox{flat universe}),\\
\sinh\chi &(k=-1, \hspace*{0.55cm}\mbox{open universe})\cdot
\end{array}\right.
\end{eqnarray}
The spacetime of the thick dust shell itself is again described by a
LTB metric (\ref{LTB}). We have to write down first the relevant
components of the (\ref{fund}). The four-velocity $u^{\mu}$ and the
normal vector $n^{\mu}$ on the spherical thick shell's core
$\Sigma_{0}$ measured by the FRW observers on both sides of the
shell are, respectively, given by
\begin{eqnarray}\label{unfrw}
u^{\mu}\Bigl|^{in/out}_{\Sigma_{0}}&=&(\dot{t},\dot{\chi},0,0)\Bigl|^{in/out}_{\Sigma_{0}},\nonumber\\
n^{\mu}\Bigl|^{in/out}_{\Sigma_{0}}&=&(a\dot{\chi},\frac{\dot{t}}{a},0,0)\Bigl|^{in/out}_{\Sigma_{0}}.
\end{eqnarray}
Dots denote again the derivative with respect to the proper time
$\tau_{0}$ on $\Sigma_{0}$. The relevant components of the Riemann
curvature tensor for the FRW spacetime are
\begin{eqnarray}\label{Riemnfrw}
R_{\theta
rr\theta}\Bigl|^{in/out}&=&-a^{2}r^{2}(\frac{d \textit{a}}{d t})^2\Bigl|^{in/out},\nonumber\\
R_{\theta tt
\theta}\Bigl|^{in/out}&=&r^{2}a\frac{d^{2}\textit{a}}{d t^2}\Bigl|^{in/out},\nonumber\\
R_{trrt}\Bigl|^{in/out}&=&-a\frac{d^{2}\textit{a}}{d
t^2}\Bigl|^{in/out}.
\end{eqnarray}
Following components of the extrinsic curvature tensor for the
interior and exterior FRW metrics evaluated on $\Sigma_{0}$ are also
required:
\begin{eqnarray}\label{kththfrw}
K_{\theta}^{\theta}\Bigl|_{\Sigma_{0}}^{in/out}&=&
\frac{\epsilon_{in/out}}{R_{0}}\sqrt{1+\dot{R_{0}}^2-\frac{8\pi
R_{0}}{3}\rho_{in/out}}\,\, ,\\
K_{\tau}^{\tau}\Bigl|^{in/out}_{\Sigma_{0}}&=&\frac{\ddot{R_{0}}+\frac{4\pi
R_{0}}{3}\rho_{in/out}+4\pi
R_{0}\rho_{in/out}\frac{v_{in/out}^{2}}{1-v_{in/
out}^{2}}}{\sqrt{1+\dot{R_{0}}^2-\frac{8\pi
R_{0}}{3}\rho_{in/out}}}\,\, .\nonumber\\
\label{kttfrw}& &
\end{eqnarray}
The standard junction, we are using, corresponds to
$(\partial\chi/\partial n)>0$. Therefore, we have
\cite{Mansouri4,Sakai}:
\begin{equation}\label{epsilonfrw}
\epsilon_{in/out}=sgn \left( \frac{d r}{d\chi} +
vHR\right)\Bigl|^{in/out}_{\Sigma_{0}},
\end{equation}
where $v_{in}$ and $v_{out}$ represent the peculiar velocities of
$\Sigma_{0}$ relative to the expansion in the interior and exterior
FRW spacetimes, respectively. These can be computed in the same way
as was done for $v_{0}$ in (\ref{upsilon}):
\begin{equation}\label{upsilonfrw}
v_{in/out}=\frac{R_{0}H_{in/out}A_{in/out}-\dot{R}_{0}}{R_{0}\dot{R}_{0}H_{in/out}-
A_{in/out}}\,\, ,
\end{equation}
where we have defined
\begin{eqnarray}\label{definitionfrw}
A_{in}&=&\sqrt{1+\dot{R_{0}}^2-\frac{8\pi
R_{0}^{2}}{3}\rho_{in}}\,\,\,\,,\nonumber\\
A_{out}&=&\sqrt{1+\dot{R_{0}}^2-\frac{8\pi
R_{0}^{2}}{3}\rho_{out}}\,\,\,.
\end{eqnarray}
Combining all these results, we get the following final form for the
angular and time component of (\ref{fund}) written up to the first
order in $w$,
\begin{eqnarray}\label{eqfrw}
\epsilon_{in}A_{in}- \epsilon_{out}A_{out}&=&\frac{8\pi
R_{0}w\rho_{0}}{1-v_{0}^2} + w\left(\frac{2M_{0}}{R_{0}^2}
-R_{0}\left(\frac{H_{in}^2}{1-v_{in}^2}+\frac{H_{out}^2}{1-v_{out}^2}\right)
\right.\nonumber\\
&-&\left.\frac{4\pi
R_{0}}{3}\left(\rho_{in}\frac{2-v_{in}^2}{1-v_{in}^2}
+\rho_{out}\frac{2-v_{out}^2}{1-v_{out}^2}\right)\right)\,\,,\\
\label{thickeq2frw} (A_{in}-A_{out})\ddot{R_{0}}&=& \frac{4\pi
R_{0}}{3}\left(A_{out}\rho_{in}\frac{1+2v_{in}^{2}}{1-v_{in}^{2}}
-A_{in}\rho_{out}\frac{1+2v_{out}^{2}}{1-v_{out}^{2}}\right)
\nonumber\\
&-&wA_{in}A_{out}\left(A_{in}^{-2}\left(\ddot{R}_{0}+\frac{4\pi}{3}R_{0}\rho_{in}
\frac{1+2v_{in}^{2}}{1-v_{in}^{2}}\right)^2\right.\nonumber\\
&+&A_{out}^{-2}\left(\ddot{R}_{0}+\frac{4\pi}{3}R_{0}\rho_{out}\frac{1+2v_{out}^{2}}
{1-v_{out}^{2}}\right)^2\nonumber\\
&-&\left.\frac{2(\ddot{R}_{0}+\frac{M_{0}}{R_{0}^2}+4\pi
R_{0}\frac{\rho_{0}v_{0}^2}{1-v_{0}^2})^2}{1+\dot{R}_{0}^2-\frac{2M_{0}}{R_{0}}}-\frac{4M_{0}}{R_{0}^3}+8\pi
\rho_{0}\right).
\end{eqnarray}
Once we know the FRW solutions in ${\cal M}_{in/out}$ (i.e.,
$H_{in/out}$ and $\rho_{in/out}$), we can solve in principle the set
of equations (\ref{eqfrw}-\ref{thickeq2frw}), (\ref{upsilonfrw}),
and (\ref{upsilon}) for the given initial conditions $R_{0}(0)$,
$\dot{R}_{0}(0)$, $E_{0}$, $F_{0}$, and constant $w$, to determine
the time evolutions of the proper radius $R_{0}(\tau_{0})$ and mass
density $\rho_{0}$ of the thick shell's core. Note that the
solutions in ${\cal M}_{in/out}$ are expressed in terms of the
cosmic times $t_{in/out}$, hence one needs to take into account the
following equation:
\begin{equation}\label{tfrw}
\frac{d t_{in/out}}{d\tau_{0}}=\frac{1}{\sqrt{1-v_{in/out}^{2}}}
\end{equation}

\subsection{Impact of thickness and the thin wall limit}

Let us now look at the effects of the thickness on the evolution of
the shell in FRW spacetimes. Using (\ref{rho}) and assuming
$v_{0}\ll1$ we rewrite (\ref{eqfrw}) in the following familiar form:
\begin{equation}\label{Israelfrw}
\epsilon_{in}\sqrt{1+ \dot{R_{0}}^{2}- \frac{8\pi
}{3}R_{0}^{2}\rho_{in}}-\epsilon_{out}\sqrt{1+\dot{R_{0}}^{2}
-\frac{8\pi}{3}R_{0}^{2}\rho_{out}}=4\pi \tilde{\sigma}R_{0},
\end{equation}
where we have again introduced an effective surface energy as
\begin{eqnarray}\label{sigmafrw}
& &\tilde{\sigma}=
\sigma\left(1+\upsilon_{0}^{2}+\frac{wM_{0}}{2\pi\sigma R_{0}^3}
\right.\nonumber\\
& &-\frac{w}{4\pi\sigma}\left(\frac{H_{in}^2}{1-v_{in}^2}
+\frac{H_{out}^2}{1-v_{out}^2}+\left.\frac{4
\pi\rho_{in}}{3}\frac{2-v_{in}^2}{1-v_{in}^2}
 +\frac{4\pi\rho_{out}}{3}\frac{2-v_{out}^2}{1-v_{out}^2}\right)\right).
\end{eqnarray}
Equation (\ref{Israelfrw}) is identical to the equation of motion of
a thin shell with the effective surface energy density
$\tilde{\sigma}$ playing the role of $\sigma$. In fact, taking the
zero thickness limit $w\rightarrow0$ and $v_{0}\rightarrow0$, we see
that (\ref{Israelfrw}) reduces to the equation of motion for a thin
shell gluing two different FRW spacetimes obtained by Maeda
\cite{Maeda} and Berezin \textit{et al}, \cite{Berezin}.
\par
Depending on the relative magnitudes of the different terms on the
right hand side of (\ref{sigmafrw}), coming from the shell itself
and the surrounding media ${\cal M}_{in/out}$, we may have
$\tilde{\sigma}<\sigma$, or vice versa. Proceeding exactly in the
same manner as before, we solve (\ref{Israelfrw}) for
$\dot{R}_{0}^{2}$ and obtain
\begin{equation}\label{Rdotfrw}
\dot{R}_{0}^{2} =4\pi^{2}  R_{0}^{2} \tilde{\sigma}^{2}
+\frac{R_{0}^2(\rho_{o}-\rho_{i})^2}{9\tilde{\sigma}^{2}}
+\frac{4\pi}{3}R_{0}^2(\rho_{o}+\rho_{i})-1\, .
\end{equation}
The zeroes of $\dot{R}_{0}^{2}$ are
\begin{equation}
\tilde{\sigma}_{\pm}^{2}
=\frac{1-\frac{4\pi}{3}R_{0}^2(\rho_{i}+\rho_{o}) \pm \sqrt{(1-
\frac{8\pi }{3}R_{0}^{2}\rho_{i})(1- \frac{8\pi
}{3}R_{0}^{2}\rho_{o})}}{8\pi^2R_{0}^2}\, . \label{Zeroes1frw}
\end{equation}
It is easily seen that in the case of
$\tilde{\sigma}>\tilde{\sigma}_{+}$ the solution (\ref{Rdotfrw}) in
which $\dot{R}_{0}^{2}$ increases with $\tilde{\sigma}$, belongs to
(\ref{Israelfrw}) with $\epsilon_{out}=-1$ (Figure 1). According to
(\ref{epsilonfrw}) this case may happen in a closed universe
$k_{out}=+1$ with $d r/d\chi<0$, and in a flat or open universe
whenever the peculiar velocity $v_{out}$ of $\Sigma_{0}$ is negative
and the radius $R_{0}$ is greater than the Hubble radius
$H_{out}^{-1}$. Then, depending on $\tilde{\sigma} < \sigma$ or
$\tilde{\sigma} > \sigma$, a comparison to the thin shell case shows
that
 the effect of the first order corrections due to the shell's thickness is to
slowdown or speed up the shell's core motion. The case
$\tilde{\sigma}_{-}<\tilde{\sigma}<\tilde{\sigma}_{+}$ is not
physical because of the relation $\dot{R}_{0}^{2}< 0$. Finally, the
case $\tilde{\sigma}<\tilde{\sigma}_{-}$ in which $\dot{R}_{0}^{2}$
is a decreasing function of $\tilde{\sigma}$, corresponds to
(\ref{Israelfrw}) with $\epsilon_{out}=+1$ implying a slowdown of
the shell for larger values of $\tilde{\sigma}$. Thus, if the outer
bracket in the (\ref{sigmafrw}) is smaller (greater) than one, then
we may say that the effect of the first order corrections to the
thin shell approximations is to speed up (down) the shell's core
motion.

\section{Conclusion}

The dynamics of a thick wall embedded in curved spacetimes has been
studied in a more general setting than our previous work
\cite{Mansouri1}. This was achieved by imposing the Darmois junction
conditions on boundaries of a codimension-1 wall with two embedding
spacetimes of any dimension. Using an expansion scheme for the
extrinsic curvature tensor of the wall boundaries in terms of its
proper thickness, we have then obtained the general equation
(\ref{fund}) describing the evolution of the wall's core  embedded
in the n-dimensional spacetime. Our treatment is valid for those
walls or branes whose thickness are small compared to their
curvature radius. We have exemplified our formalism in two different
cases of walls immersed in vacuum and walls immersed in FRW
spacetimes. Taking the spacetime within the wall itself to be
inhomogeneous of the LTB-type, in both cases the peculiar velocity
of the wall's core relative to the LTB metric of the wall turns out
to play an important role.\\
We have written down the equations of motion for the wall, up to the
first order of the proper thickness, and it was shown in both cases
that in the limit of the thin wall we obtain the known results of
the corresponding thin walls. In fact the equations of motion could
be written in a form similar to the thin wall cases with an
effective surface density. In the case of a wall immersed in vacuum,
it turns out that the effect of the proper thickness $w$ is to speed
up the collapse of the comoving spherical dust shell, while the
first order peculiar velocity correction tends to decrease the
collapse velocity. In the case of a dust thick shell immersed in FRW
spacetimes, the first order effect of the thickness is much more
complicated, depending on different parameters coming from the FRW
spacetimes on both sides of the shell and the LTB spacetime of the
shell itself.\\
The formalism presented is quite general. It can be applied to any
other type of the wall or brane, such as gravitating domain walls in
curved spacetime having a planar or spherical symmetry or thick
branes in brane cosmology, given that the geometry of the wall
itself is known.
\section{Acknowledgments}
R.M. would like to thank McGill Physics department, specially Robert
Brandenberger, for the hospitality. Sh. Khosravi wishes to thank Dr
Sima Ghassemi for useful discussions. This work was partially
supported by TWASIC( TWAS Iran chapter) through ISMO, Tehran.

\end{document}